\documentclass[12pt]{article}
\usepackage{epsfig}
\usepackage{float}
\usepackage{amssymb,amsmath}
\newcommand{\be}{\begin{equation}}
\newcommand{\ee}{\end{equation}}
\newcommand{\bea}{\begin{eqnarray}}
\newcommand{\eea}{\end{eqnarray}}
\begin{document}
\begin{center}
 {\bf \begin{Large}{M\"ossbauer antineutrinos: some basic considerations}\end{Large}}\footnote{Based on a lecture presented at the XXXIII International Conference of Theoretical Physics, MATTER TO THE DEEPEST: Recent Developments in Physics of Fundamental Interactions, Ustro\'n, Poland, September 11-16, 2009. \textit{Published in ACTA PHYSICA POLONICA \textbf{B40}, 3033 (2009).}}
\end{center}
\begin{center}
\bf {Walter Potzel}
\end{center}
\begin{center}
{\em Physik-Department E15, Technische Universit\"at M\"unchen 
\\D-85748 Garching, Germany}
\end{center}

\begin{abstract}

Basic aspects of phononless resonant capture of monoenergetic electron antineutrinos (M\"ossbauer antineutrinos) emitted in bound-state $\beta$-decay in the $^{3}$H - $^{3}$He system are considered. It is shown that stochastic magnetic relaxation phenomena as well as the direct influence of solid-state effects on the energy of the $\bar{\nu_{e}}$ will cause line broadening by a factor of more than $10^{13}$. Lattice expansion and contraction after the transformation of the nucleus will drastically reduce the probability for phononless transitions. Thus, the observation of M\"ossbauer $\bar{\nu_{e}}$ of the $^{3}$H - $^{3}$He system will most probably be unsuccessful. As a possible alternative, the Rare-Earth system $^{163}$Ho - $^{163}$Dy is briefly discussed.
\end{abstract}

PACS numbers: 14.60.Pq, 13.15.+g, 76.80.+y, 76.20.+q

\section{Introduction}

After the discovery of the M\"ossbauer effect with $\gamma$-transitions in nuclei \cite{Moessbauer}, the conditions for observing M\"ossbauer antineutrinos, i.e., the recoilfree resonant emission and absorption of electron antineutrinos ($\bar{\nu_{e}}$), have been discussed in various publications \cite{Visscher},\cite{Kells},\cite{RajuRag},\cite{WalterPotzel},\cite{Raghavan1}. The basic concept is to use 18.6 keV $\bar{\nu_{e}}$ that are emitted in the bound-state $\beta$-decay of $^{3}$H to $^{3}$He \cite{Bahcall},\cite{Mika} and are resonantly captured in the reverse bound-state process in which $^{3}$He is transformed to $^{3}$H \cite{Kells},\cite{RajuRag},\cite{WalterPotzel},\cite{WalterPotzel1}. Since bound-state $\beta$-decay  is a two-body process, the emitted $\bar{\nu_{e}}$ has a fixed energy $E_{\bar{\nu_{e}}}=Q+B_{z}-E_{R}$. The $\bar{\nu_{e}}$-energy is determined by the Q value, the binding energy $B_{z}$ of the atomic orbit the electron is emitted into, and by the recoil energy $E_{R}$ of the atom formed after the decay. Also the reverse reaction is a two-body process. The required $\bar{\nu_{e}}$-energy is given by $E'_{\bar{\nu_{e}}}=Q+B_{z}+E'_{R}$, where $E'_{R}$ is the recoil energy of the atom after the transformation of the nucleus. To prevent the recoil, $^{3}$H as well as $^{3}$He are considered to be imbedded in Nb metal lattices.

In analogy to conventional M\"ossbauer spectroscopy with photons, the absorption cross section $\sigma_R$ at resonance for M\"ossbauer antineutrinos can be written as \cite{Moessbauer},\cite{Kells}

\begin{equation}\label{cross section}
\sigma_{R}=2\pi\left(\frac{\lambda}{2\pi}\right)^{2} s^{2} (\alpha^{2} f^{2} \delta) ~,
\end{equation}
where $\lambda$ is the wavelength of the antineutrinos, $s$ takes into account statistical factors (nuclear spins, isotopic abundance, etc.) and is considered to be of the order of unity, $\alpha$ is the ratio of bound-state to continuum-state $\beta$ decays ($\alpha\approx0.005$ for the $^{3}$H - $^{3}$He system), $f$ is the probability that no phonons are excited in the Nb lattice when the $\bar{\nu_{e}}$ is emitted or captured, and $\delta=\Gamma/\Gamma_{exp}$, with $\Gamma$ the natural linewidth of the weak decay and $\Gamma_{exp}$ the experimental width due to line broadening. If $f$ and $\delta$ would be of the order of unity,

\begin{equation}\label{cross section ideal}
\sigma_{R}\approx1.8\cdot10^{-22} cm^{2}
\end{equation}
for $\bar{\nu_{e}}$ with an energy of 18.6 keV ($\lambda\approx0.67\cdot10^{-8}$cm). This would be a huge cross section for all standards of neutrino interactions.

In the present paper, we will demonstrate that - in contradiction to a recent publication \cite{RaghavanPRL} - it will not be possible to observe $\Gamma=h/{(2\pi\tau)}=1.17\cdot10^{-24}eV$, $\tau=17.81y$ being the lifetime of $^{3}$H. Magnetic relaxation phenomena in Nb metal as well as the {\em direct} influence of the variation of the binding energies (due to the random distributions of $^{3}$H and $^{3}$He in the Nb lattices) on the energy of the $\bar{\nu_{e}}$ will cause $\Gamma_{exp}\gg\Gamma$ by many orders of magnitude. In addition, the factor $f^2$ in Eq.(\ref{cross section}) may be as small as $\sim 10^{-7}$ due to lattice expansion and contraction processes which - being not present in conventional M\"ossbauer spectroscopy with photons - have been overlooked until recently \cite{WalterPotzel1}, \cite{WalterPotzelPRL}. With the ideas and techniques available at present, the observation of $\bar{\nu_{e}}$ will most probably be unsuccessful. We briefly consider the Rare-Earth system $^{163}$Ho - $^{163}$Dy as an alternative.

\section{Phononless transitions}

Concerning $\bar{\nu_{e}}$, phononless transitions are jeopardized by two kinds of lattice-ex\-cit\-ation processes: 
\begin{enumerate}
\item momentum transfer due to emission/capture of a $\bar{\nu_{e}}$,
\item lattice expansion and contraction when the nuclear transformation occurs during which the $\bar{\nu_{e}}$ is emitted or absorbed. This possibility is not present in usual M\"ossbauer spectroscopy with photons \cite{WalterPotzel1},\cite{WalterPotzelPRL}.
\end{enumerate}

Considering first the momentum transfer, in the Debye-model and in the limit of very low temperatures $T$, the recoilfree fraction $f_r$ is given by

\begin{equation}\label{recoilfree fraction}
f_{r}(T\rightarrow0)=exp \left\{-\frac{E^{2}}{2Mc^{2}}\cdot\frac{3}{2k_{B}\theta}\right\},
\end{equation}
where $\theta$ is the effective Debye temperature, $k_{B}$ is the Boltzmann constant, and $\frac{E^{2}}{2Mc^{2}}$ is the recoil energy which would be transmitted to a free atom of mass $M$. With $\theta\approx800K$ \cite{WalterPotzel1}, $f_{r}^2\approx0.07$ for $T\rightarrow0$.

Considering process 2, the $\bar{\nu_{e}}$ itself takes part in the nuclear processes transforming one chemical element into a different one. $^{3}$H and $^{3}$He are differently bound in the Nb lattice and use different amounts of lattice space. Thus, when the nuclear processes occur, the lattice will expand or contract. The lattice deformation energies for $^{3}$H and $^{3}$He in the Nb lattice are $E_L^{^{3}H}=0.099$ eV and $E_L^{^{3}He}=0.551$ eV, respectively \cite{Puska}. Assuming again an effective Debye temperature of $\theta\approx800$ K one can estimate - in analogy to the situation with momentum transfer - that the probability $f_L$ that this lattice deformation will {\em not} cause lattice excitations is smaller than

\begin{equation}\label{lattice deformation}
f_L\approx exp \left\{-\frac{E_L^{^{3}He}-E_L^{^{3}H}}{k_{B}\theta}\right\}\approx1\cdot10^{-3}.
\end{equation}
Thus, the total probability for phononless emission and consecutive pho\-non\-less capture of $\bar{\nu_{e}}$ is
\begin{equation}\label{phononless probability}
f^2=f_r^2\cdot f_L^2\approx7\cdot10^{-8},
\end{equation}
 a very tiny probability indeed.

\section{Linewidth}

Basically, there are three important types of line-broadening effects \cite{Balko},\cite{Coussement}, \cite{WalterPotzeletal}, which prevent the observation of the natural width of $\bar{\nu_{e}}$ in the system $^{3}$H - $^{3}$He, imbedded in Nb metal.

\subsection{Homogeneous broadening}
Homogeneous broadening is caused by electromagnetic relaxation, e.g., by spin-spin interactions between the nuclear spins of $^{3}$H and $^{3}$He and with the spins of the Nb nuclei. Contrary to the claim in Ref. \cite{RaghavanPRL}, such magnetic relaxations are stochastic processes and can \textit{not} be described by a periodic energy modulation of an excited hyperfine state \cite{WalterPotzel1},\cite{WalterPotzelPRL}. Stochastic processes lead to sudden, irregular transitions between hyperfine-split states originating from many magnetic spins. Thus the wave function of the $\bar{\nu_{e}}$ is determined by random phase changes which lead to line broadening characterized by the time-energy uncertainty relation \cite{WalterPotzel1}. As a consequence, the broadened lines can \textit{not} be decomposed into multiple sharp lines. For the system $^{3}$H - $^{3}$He in Nb metal, $\Gamma_{exp}\approx4\cdot10^{13}\Gamma$, thus $\delta_H\approx2.5\cdot10^{-14}$ \cite{WalterPotzel1},\cite{WalterPotzelPRL}.

\subsection{Inhomogeneous broadening}
This type of broadening is caused by stationary effects, e.g., by lattice defects, impurities and, in particular, by the random distributions of $^{3}$H and $^{3}$He on the tetrahedral interstitial sites in Nb metal. Thus the periodicity of the lattice is destroyed and the binding energies $E_B$ of $^{3}$H and $^{3}$He will vary within the Nb lattice. Typical values for $E_B$ are in the eV range \cite{Puska}. In conventional M\"ossbauer spectroscopy with photons, using the best single crystals, such variations in $E_B$ cause energy shifts of $\approx10^{-12}eV$. Since, in the nuclear transformations, the $\bar{\nu_{e}}$ energy is \textit{directly} affected by $E_B$, one has to expect that the variations of the $\bar{\nu_{e}}$ energy are much larger than $10^{-12}eV$, probably in the $10^{-6}eV$ regime. Thus, inhomogeneous line broadening is estimated to give $\delta_I\ll10^{-12}$ \cite{WalterPotzel1},\cite{WalterPotzelPRL}, probably $\delta_I\approx10^{-18}$.

\subsection{Relativistic effects}
Another contribution to line broadening in an imperfect lattice is due to relativistic effects. An atom vibrating around its equilibrium position in a lattice exhibits a mean-square velocity $\left\langle v^{2}\right\rangle$. According to Special Relativity Theory this causes a time-dilatation, i.e., a reduction in frequency (energy $E$) of the $\bar{\nu_{e}}$: $\Delta E=-v^{2}E/(2c^{2})$. Being proportional to $(v/c)^{2}$, this reduction is often called second-order Doppler shift (SOD).

Within the Debye model, the SOD between source (s) at temperature $T_{s}$ and target (t) at temperature $T_{t}$ is given by

\begin{equation}\label{SOD}
(\Delta E/E)=\frac{9k_{B}}{16Mc^{2}}(\theta_{s}-\theta_{t})
+\frac{3k_{B}}{2Mc^{2}}\left[T_{s}\cdot f(T_{s}/\theta_{s})-T_{t}\cdot f(T_{t}/\theta_{t})\right]
\end{equation}

with
\begin{equation}\label{Debye integral}
f(T/\theta)=3\left(\frac{T}{\theta}\right)^{3}\cdot\int^{\theta/T}_{0}\frac{x^{3}}{exp(x)-1}dx.
\end{equation}

At low temperatures (e.g., in a liquid-He bath at 4.2 K), the tem\-per\-a\-ture-dependent term in Eq.(\ref{SOD}) can usually be neglected if source and target are at about the same temperature. However, even in the low-temperature limit, the first term in Eq.(\ref{SOD})which is caused by the zero-point motion (zero-point energy) can not be neglected. It is {\em not} required that the chemical bonds (i.e. the Debye temperatures) of $^{3}$H and $^{3}$He have to be the same in the metal matrix. However, the chemical bond for $^{3}$H (and also for $^{3}$He) has to be the same in source and target \cite{WalterPotzel}. The critical point is, that a variation of the binding energies $E_B$ of $^{3}$H and $^{3}$He in an imperfect lattice will result in a variation of the effective Debye temperatures and thus also in a variation of the zero-point energies. If the effective Debye temperature varies by only 1 K, $(\Delta E/E)\approx2\times10^{-14}$, corresponding to a lineshift of $3\times10^{14}$ times the natural width $\Gamma$. Thus, this broadening effect is expected to give $\delta_{SOD}\approx3\cdot10^{-15}$ comparable to or even smaller than $\delta_H$ \cite{WalterPotzel1}.

\bigskip
With $\lambda\approx0.67\cdot10^{-8}$cm for the wavelength of the $\bar{\nu_{e}}$, $\alpha^2\approx2.5\cdot10^{-5}$, $f^2\approx7\cdot10^{-8}$ and $\delta_H\leq2.5\cdot10^{-14}$, we get from Eq.(\ref{cross section}) an optimistic upper limit of
\begin{equation}\label{upper limit}
\sigma_R\leq3\cdot10^{-43} cm^2.
\end{equation}

Thus, in particular the small values for $\delta$ (various types of broadening) make it impossible to reach the natural width and together with the small value for $f$ (probability for phononless transitions) it appears that M\"ossbauer $\bar{\nu_{e}}$ can not be observed with the system $^{3}$H - $^{3}$He, imbedded in Nb metal. We have reached this conclusion by only taking into account some basic principles. Technological difficulties, which are expected to be enormous too \cite{Schiffer}, have not been considered in the present paper.

\section{Rare-Earth systems - an alternative?}

In Ref. \cite{Kells}, the system $^{163}$Ho - $^{163}$Dy was estimated to be the next-best case. Compared to $^{3}$H - $^{3}$He, there are three major advantages:
\begin{enumerate}

\item The Q value of 2.6 keV (thus the $\bar{\nu_{e}}$ energy) is very low, the mass of the nuclei is large. As a consequence, the recoilfree fraction is expected to be $\approx1$.

\item Due to the similar chemical behaviour of the Rare Earths, the lattice deformation energies of $^{163}$Ho and $^{163}$Dy can be expected to be similar and could lead to a probability of phononless transitions which is larger by up to seven orders of magnitude.

\item Due to the large mass, the relativistic effects mentioned in Section 3.3, will be smaller.
\end{enumerate}

The main disadvantage will be the large magnetic moments due to the 4f electrons of the Rare-Earth atoms. Thus, line broadening effects due to magnetic relaxation phenomena will again be decisive for a successful observation of M\"ossbauer $\bar{\nu_{e}}$. Profound technological knowledge concerning the fabrication of high-purity materials and single crystals in large amounts will be required \cite{Kells}. Further investigations are necessary.

\section{Interesting experiments}
Several basic questions and interesting experiments could be addressed if M\"ossbauer $\bar{\nu_{e}}$ could be observed \cite{WalterPotzel1}:

\begin{enumerate}
\item M\"ossbauer $\bar{\nu_{e}}$ could lead to a better understanding of the true nature of neutrino oscillations\cite{Bilenkyetal1},\cite{Bilenkyetal2},\cite{Akhmedov},\cite{Akhmedov1},\cite{Bilenkyetal3},\cite{Bilenkyetal4},\cite{Kopp}. Considering the evolution of the neutrino state in time only, neutrino oscillations are a non-stationary phenomenon and M\"ossbauer $\bar{\nu_{e}}$ would not oscillate because of their extremely narrow energy distribution. An evolution of the neutrino wave function in space and time, however, would make oscillations possible in both the non-stationary and also in the stationary case (M\"ossbauer $\bar{\nu_{e}}$).
\item If M\"ossbauer $\bar{\nu_{e}}$ oscillate, the low energy (18.6 keV) would allow us to use ultra-short base lines ($\approx10$ m instead of 1500 m) to determine accurate oscillation parameters, e.g., $\Theta_{13}$, $\Delta m^{2}_{12}$ and $\Delta m^{2}_{31}$ \cite{Minakata}. The question of mass hierarchy could be settled: for normal (inverted) hierarchy, the phase of atmospheric-neutrino oscillations advances (is retarded) by $2\pi\cdot\sin^{2}\Theta_{12}$ for every solar oscillation \cite{Minakata1},\cite{Parke}.
\item Oscillating M\"ossbauer $\bar{\nu_{e}}$ could be used to search for the conversion $\bar{\nu_{e}}$ $\rightarrow\nu_{sterile}$ \cite{Kopeikin}. For $\Delta m^{2}\approx1$ eV, the oscillation length would only be $\sim5$ cm!
\item Gravitational redshift $\bar{\nu_{e}}$ measurements could only be performed if an experimental linewidth of $\Gamma_{exp}\approx10^{-10}$eV (or smaller) could be reached \cite{WalterPotzel1}. Unfortunately, our estimates in this paper show that with the $^{3}$H - $^{3}$He system, this is unrealistic.

\end{enumerate}

\section{Conclusions}

Several basic aspects of the $^{3}$H - $^{3}$He system have been considered for a possible observation of M\"ossbauer $\bar{\nu_{e}}$. In contradiction to the claim of Ref. \cite{RaghavanPRL} it will not be possible to reach the natural linewidth because of homogeneous broadening - due to stochastic relaxation processes - and inhomogenous broadening mainly due to the variations of binding energies and zero-point energies and their \textit{direct} influence on the energy of the $\bar{\nu_{e}}$. In addition, the probability for phononless emission and detection will drastically be reduced due to lattice expansion and contraction after the transformation of the nucleus. The observation of M\"ossbauer $\bar{\nu_{e}}$ of the system $^{3}$H - $^{3}$He, imbedded in Nb metal, will most probably \textit{not} be possible.
The Rare-Earth system $^{163}$Ho - $^{163}$Dy offers several advantages, in particular a large probability of phononless emission and detection. However, magnetic relaxation processes and technological requirements still have to be investigated in detail to decide if the $^{163}$Ho - $^{163}$Dy system is a successful alternative. Interesting experiments could be performed if M\"ossbauer $\bar{\nu_{e}}$ were observed.

\bigskip
\textbf{Acknowledgments}\\
This work was supported by funds of the Deutsche Forschungsgemeinschaft DFG (Transregio 27: Neutrinos and Beyond), the Munich Cluster of Excellence (Origin and Structure of the Universe), and the  Maier-Leibnitz-Laboratorium (Garching).


\begin{thebibliography}{99}

\bibitem{Moessbauer} R.L.M. M\"ossbauer, Z. Physik \textbf{151}, 124 (1958).
\bibitem{Visscher} W.M. Visscher, Phys. Rev. \textbf{116}, 1581 (1959).
\bibitem{Kells} W.P. Kells and J.P. Schiffer, Phys. Rev. C\textbf{28}, 2162 (1983).
\bibitem{RajuRag} R.S. Raghavan, hep-ph/0601079 v3, (2006).
\bibitem{WalterPotzel} W. Potzel, Phys. Scr. \textbf{T127}, 85 (2006).
\bibitem{Raghavan1} R.S. Raghavan, arXiv: 0805.4155 [hep-ph] and 0806.0839 [hep-ph] (2008).
\bibitem{Bahcall} J.N. Bahcall, Phys. Rev. \textbf{124}, 495 (1961).
\bibitem{Mika} L.A. Mika\'elyan, B.G. Tsinoev, and A.A. Borovoi, Sov. J. Nucl. Phys.  \textbf{6}, 254 (1968).
\bibitem{WalterPotzel1} W. Potzel, J. Phys.: Conf. Ser. \textbf{136}, 022010 (2008), arXiv: 0810.2170 [hep-ph].
\bibitem{RaghavanPRL} R.S. Raghavan, Phys. Rev. Lett. \textbf{102}, 091804 (2009).
\bibitem{WalterPotzelPRL} W. Potzel, Phys. Rev. Lett. \textbf{103}, 099101 (2009), arXiv: 0908.3985 [hep-ph].
\bibitem{Puska} M.J. Puska and R.M. Nieminen, Phys. Rev. B\textbf{10}, 5382 (1984).
\bibitem{Balko} B. Balko, I.W. Kay, J. Nicoll, and J.D. Silk, Hyperfine Interact.  \textbf{107}, 283 (1997).
\bibitem{Coussement} R. Coussement, G. S'heeren, M. Van Den Bergh, and P. Boolchand, Phys. Rev. B\textbf{45}, 9755 (1992).
\bibitem{WalterPotzeletal} W. Potzel et al., Hyperfine Interact. \textbf{72}, 197 (1992).
\bibitem{Schiffer} J.P. Schiffer, Phys. Rev. Lett. \textbf{103}, 099102 (2009).
\bibitem{Bilenkyetal1} S.M. Bilenky, F. von Feilitzsch, and W. Potzel, Phys. Part. Nucl. \textbf{38}, 117 (2007); and J. Phys. G: Nucl. Part. Phys. \textbf{34}, 987 (2007).
\bibitem{Bilenkyetal2} S.M. Bilenky, F. von Feilitzsch, and W. Potzel, J. Phys. G: Nucl. Part. Phys. \textbf{35}, 095003 (2008), arXiv: 0803.0527 v2 [hep-ph].
\bibitem{Akhmedov} E.Kh. Akhmedov, J. Kopp, and M. Lindner, J. High Energy Phys. \textbf{0805}, 005 (2008), arXiv: 0802.2513 [hep-ph].
\bibitem{Akhmedov1} E.Kh. Akhmedov, J. Kopp, and M. Lindner, J. Phys. G: Nucl. Part. Phys. \textbf{36}, 078001 (2009), arXiv: 0803.1424v2 [hep-ph].
\bibitem{Bilenkyetal3} S.M. Bilenky, F. von Feilitzsch, and W. Potzel, J. Phys. G: Nucl. Part. Phys. \textbf{36}, 078002 (2009), arXiv: 0804.3409 [hep-ph].
\bibitem{Bilenkyetal4} S.M. Bilenky, F. von Feilitzsch, and W. Potzel, arXiv: 0903.5234 [hep-ph], (2009).
\bibitem{Kopp} J. Kopp, J. High Energy Phys. \textbf{0906}, 049 (2009), arXiv: 0904.4346 [hep-ph].
\bibitem{Minakata} H. Minakata and S. Uchinami, New J. Phys. \textbf{8}, 143 (2006), hep-ph/0602046.
\bibitem{Minakata1} H. Minakata, H. Nunokawa, S.J. Parke, and R. Zukanovich Funchal, Phys. Rev. D\textbf{76}, 053004 (2007), arXiv: hep-ph/0701151.
\bibitem{Parke} S.J. Parke, H. Minakata, H. Nunokawa, and R. Zukanovich Funchal, Nucl. Phys. Proc. Suppl. \textbf{188}, 115 (2008), arXiv: 0812.1879 [hep-ph].
\bibitem{Kopeikin} V. Kopeikin, L. Mikaelyan, and V. Sinev, hep-ph/0310246v2 (2003).



\end{thebibliography}
\end{document}